# Lightweight Container-based User Environment

Wenzhe Zhang, Kai Lu, Ruibo Wang, Wanqing Chi, Mingtian Shao, Huijun Wu
*College of Computer*
*National University of Defense Technology*
Changsha, China
{zhangwenzhe, kailu, ruibo, chiwq}@nudt.edu.cn
s1437252602@gmail.com

whj_nudt@foxmail.com

Mikel Luján
*School of Computer Science*
*The University of Manchester*

Manchester, UK
mikel.lujan@manchester.ac.uk

Xiaoping Wang
*College of Computer Science and Electronic Engineering*
*Hunan University*
Changsha, China
xiaopingwang@nudt.edu.cn

*Abstract*—Modern operating systems all support multi-users that users could share a computer simultaneously and not affect each other. However, there are some limitations. For example, privacy problem exists that users are visible to each other in terms of running processes and files. Moreover, users have little freedom to customize the system environment. Last, it is a burden for system administrator to safely manage and update system environment while satisfying multiple users.

Facing the above problems, this paper proposes CUE, a Lightweight Container-based User Environment. CUE proposes a new notion that stands in between application container and operating system container: user container. CUE is able to give users more flexibility to customize their environment, achieve privacy isolation, and make system update easier and safer. Its goal is to optimize and enhance the multi-user notion of current operating system and being lightweight. Moreover, it is able to facilitate application deployment in high performance clusters. It is currently deployed in NUDT's Tianhe E prototype supercomputer. Experiment results show that it introduces negligible overhead.

## I. Introduction

Modern operating systems are all multi-user systems as to enable multiple people to use seamlessly a single computer. In this scenario, there are two types of users: system administrator (root) and normal users (non-root users). Root user has full control of the system and is mainly responsible for maintaining the system environment on which non-root users run their daily jobs. In this sharing model, some shortcomings can be identified:

(1) Privacy: Users are visible to each other in terms of running processes and files. For example, a user can see what others are running (using the ps command) and the files of others. When multi-users are sharing a server, users may not want this privacy exposure to happen.

(2) Standard environment: All non-root users have little scope to customize the predefined system environment. Think about the following scenarios: "What if we have an application that links to a special library which has to be located in /lib or /usr/lib?", "What if we want a specific glic version?", "What if we need to change the configure file (located in /etc) for ssh?". These simply things cannot be done with the current sharing model.

(3) Unsafe system upgrade: When the root user wants to upgrade the system environment (for example, upgrade several libraries), the process may cause system crash or become unstable, making a set of applications and services not behave as before. Often, much effort is engaged to understand, diagnose, and then recover to the stable system environment.

The described shortcomings are well known and we have plenty of tools to deal with parts of them, but not a coherent single approach supported by a single framework. For example, application container[1] (such as docker[2-5]) addresses the problem of applications relying on various libraries. It packs applications along with their dependencies (libraries) and focuses on facilitating application deployment. In this model, users are building and running application containers instead of bare applications. However, for the users themselves (non-root users), they still have no right to customize the system environment they work/develop on. Moreover, the problems of privacy among users and unsafe system upgrade are not covered by this approach.

It may be argued that we could build a larger container that contains the whole system and users can do everything inside their own containers, isolated from each other. This takes us to another approach: Operating system container (such as LXC [4], OpenVZ [6-7]) or virtual machine [8-13]. They give each user a whole private system abstraction which is sometimes too heavyweight in terms of performance and management overhead. On one hand, these approaches usually introduce noticeable performance overhead. On another hand, they introduce more management overhead as they have changed the multi-user sharing concept. Every user now is the administrator of its own system (virtual machine or operating system container) and the root user is now responsible for managing these virtual machines. The virtual machines are black boxes to the root user and it is hard for the root user to help manage the environment inside each virtual machine.

Above all, to share a computer, we often do not need virtual machines or application containers, we just need a more isolated and more flexible environment to enhance the multi-user sharing model.

This paper proposes a new container framework situated between application container and operating system container (as well as virtual machine): user container. Based on the currently maturing container-related support in Linux kernel (e.g. namespace, cgroups, overlay file system) [14-20], we are able to provide a Lightweight Container-based User Environment (CUE) that solves the described shortcomings. In detail, CUE enables the following functionalities:

(1) When non-root users login on a server, they automatically login into special containers that represent their own user environment. The environments are the same as the standard environment that the root user manages. Non-root Users can become root in their own user environments. From their point of view, they can see and modify (almost) all the files, which gives them much flexibility for customizing their own environments. Moreover, they cannot see the processes or files of other users'. All modifications will not be seen by other users, which improves privacy.

(2) When the root user wants to update the system environment, he just does so. All users in their container will see the update immediately. For example, if the root user adds in new library in /lib, all non-root users will see the library immediately, even when they are isolated in different user environments.

(3) When a non-root user does a customization, for example, replace a library with another special version, this customization has the highest priority. That is to say, after that if the root user updates the system environment and replaces the very library with a new version, the non-root user still sees his customized one. All other users see the updated one.

(4) To achieve safe system update, the root user can also open a new user container, which has the same environment with the host at creation time. Then the root user can do all the update, test them well, and then do the update to the host environment.

In summary, CUE gives users more flexibility to customize their environment, enables privacy isolation, and makes system update easier. It is currently used in NUDT's (National University of Defense Technology) high performance computers (Tianhe E prototype system) where several users are sharing one login node to do development and job allocation. The goal of CUE is to optimize and enhance the multi-user notion of current operating system and being lightweight. We have opensourced it at https://github.com/zhangwenzhe/NUDT-cue and hope it would be adopted widely.

The rest of this paper is organized as follows: Section 2 gives the background and discussion of related work. We introduce the design and implementation in Section 3. Section 4 gives and discusses typical use cases of CUE. Section 5 gives the experimental result and we conclude in Section 6.

## II. BACKGROUND AND RELATED WORK

Virtualization [8] has a long history and has been studied extensively. It facilitates sharing, configuring, resource management, and many other aspects. Container [4] is a type of virtualization. Its history could be traced back to "chroot" [21] in UNIX, which was introduced to facilitate multi-user isolation and processes configuration. After that, "jail" from FreeBSD enables a more isolated model. At the same time, virtual machine (e.g. Xen, KVM [22]) was developing fast and widely adopted commercially, especially on cloud computing deployments. However, a key disadvantage of virtual machine is the performance overhead. As an alternative, OpenVZ [7] proposed a kind of operating system virtualization that could replace virtual machine in many cases and be much faster than traditional virtual machine. We refer to it as "operating system container" although at that time (around 2000-2005) the word "container" had not been used widely. It isolates a group of processes from many aspects to let the processes have the illusion of owning their own operating system. Realizing this needs much support from operating system kernel. OpenVZ made specific modifications to the Linux kernel to realize that and such modifications of Linux kernel were learnt by Linux group and partially adopted, which is currently the standard kernel support for modern container technologies. LXC (Linux Container) [23-24], based on the Linux namespaces, was proposed later. Like OpenVZ, LXC proposes the same notion of OS container, which is mainly meant to provide (almost) the same functionality that virtual machines do, but at lower performance cost. Until docker, the mainstream usage of container technology was not new, but just focused on providing a faster alternative to virtual machines. Unlike previous OS containers, Docker proposed a new notion called application container, which packs applications with all their dependences to facilitate the application deployment. Docker is also based on the existing Linux kernel support (namespace, cgroups, overlay file systems).

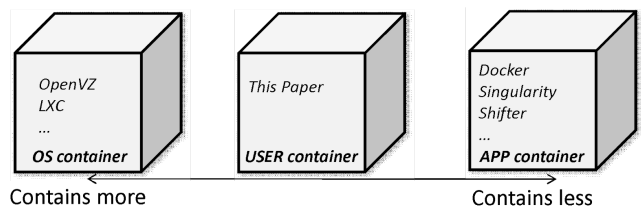

Fig. 1. Notion of user container

Above all, currently, talking about container, there are two mainstream models or notions: OS container and application container. OS container is like a virtual machine and gives a whole isolation. Application container is more like a packing tool. They focus on different scenarios and both are quite successful. In this paper, we propose a new notion that stands in between OS container and application container: User container (Figure 1). Based on currently maturing container support in Linux kernel, we aim to enhance and optimize the traditional multi-user environment of operating system, enabling better privacy among users, more flexible system environment, and safe management and upgrade of system environment. We try to realize the above features and at the same time not to change the traditional notion of multi-user.

There are also studies to embrace container technology in HPC (High Performance Computing) world. Shifter[25-26] and Singularity [27] are typical ones. They are both application containers and focus on application deployment on high performance clusters. Unlike them, our CUE focuses on enabling a more isolated and flexible development environment for multi-users to share a server. As we will show later as a use case, our CUE could also facilitate application deployment on high performance clusters.

### III. CUE - DESIGN AND IMPLEMENTATION

This section describes our design choices as well as the implementation of CUE. In short, in a user container, what a user can see and do. The source code is at https://github.com/zhangwenzhe/NUDT-cue.

#### A. Isolation and Privacy

When multi-users are sharing a server, users may want several things to be properly isolated. The spirit is, a user's behavior should only be seen by himself and not to affect others'.

1. File system

In current multi-user model, the isolation mainly happens on the file system. Modern file systems are pretty mature that different controls (e.g. access rules) are applied to different files. However, there are two main drawbacks:

(1) Non-root users cannot modify the common public files, e.g. libraries in /lib and configure files in /etc, which may be a limitation when non-root users want to customize the system environment.

(2) If a non-root user makes a modification to the public system environment via "sudo", e.g. change Python 2.7 to 3.5, all other users will see the modification. It is not uncommon to have experienced a user makes a mistake while changing the system environment (common libraries, configure files and so on), which renders the systems unusable for other multi-users sharing the server. Thus, an ideal case is that if a user makes any modification, the modification should only be seen by that user, not by any others, while if the root makes a change to the system environment, the change should be seen by all.

To solve the above problems, CUE gives each user a whole system environment. In other words, every user has its own files plus all public files (the files owned by root). Thus, users can make any customization of system environment as then need. Moreover, whenever they make any modification, the modification is only visible to themselves. CUE achieves this based on the overlay file system. The design is shown in Figure 2.

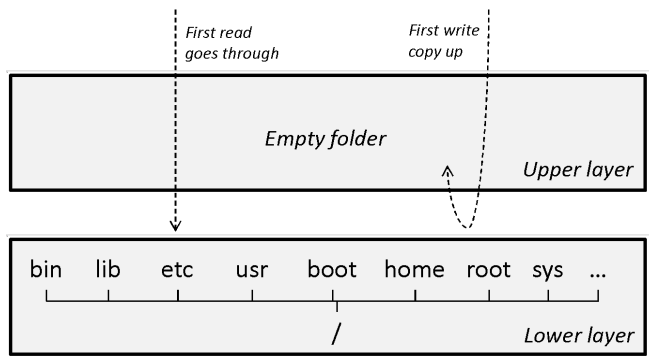

Fig. 2. CUE's file system isolation design

For each user, CUE creates a user container by overlaying an empty folder on the host "/" (shown in Figure 2) and chroot into the merged folder. Thus, from the point of view of each user, he sees the system environment the same as host. For all the public files owned by root, CUE gives the user root privilege to read and modify them. This enables users to customize their own system environment. For example, the user can change the glibc to another special version. Moreover, when he is doing this, the overlay system will put the new glibc in the upper layer, leaving the lower layer file system unchanged. This achieves the isolation that a user's modification to the file system is only seen by himself, not by others. Above all, CUE gives non-root users more freedom to customize the system files and at the same time improves privacy among users.

2. Process

A user should only see the processes he runs. All other processes should not be visible to that user. CUE uses Linux's pid namespace and remounts the proc file system to achieve process isolation for user container.

3. Hostname

As CUE allows users to do their customization on system files and configuration files, we use Linux UTS namespace to achieve hostname isolation. Users can customize the hostname.

4. Device and Network

In the traditional multi-user model, devices are shared by all users. CUE does the same and not to isolate devices. Unlike virtual machine that virtualizes devices for each virtual machine, CUE's purpose is to enhance multi-user sharing model, not to give them the illusion that the computer is exclusively owned by one. Moreover, device virtualization is too heavy weight and may incur performance cost, especially for high performance computing. Thus, CUE simply exposes the host devices to the user container. For the same reason, CUE dose not virtualizes the network devices nor uses network namespace in Linux kernel.

#### B. System Environment Management and Customization

As mentioned in section 3.1.1, CUE actually gives each user a whole system environment and proper privileges to make modification. Thus, they can make any customization and not to affect others. This is much like virtual machine. However, the difference is, it is easy for the root to manage the system environment for all users, not like the virtual machine scenario where the root is mainly responsible for managing virtual machines which are like block boxes.

As shown in Figure 2, for all non-root users, CUE gives them an overlayed folder as their root directory to ensure isolation, while for root, he just sees and manages the lower layer. Thus, if he makes any modification to the lower layer, all users will see the modification. For example, if root user changes the /bin/python from version 2.7 to 3.5, all users will see the update. Furthermore, if a user likes to use Python 2.7, he is free to change it back to 2.7 without interfering others. Finally, once the user makes such customization, further update of this file by root will not be visible to the user. This is because the file is now located in the upper layer in the layered file system and further changes of the file in the lower layer will be masked. The spirit is, users are enabled to customize everything, and once they did, the customized version has the highest priority. For the non-customized files, it is the root's job to manage and update them.

*C. Security and Capabilities*

In order to give unprivileged users more flexibility to customize the system environment, CUE gives them part of root privileges inside their own user containers. Each user logins into the user container as root. CUE relies on Linux capabilities to make fine-grained confine of what they can do.

Users are enabled to bypass file and directory permission checks thus they can modify any files they want. As introduced before, we use overlay file system to achieve isolation. Arbitrary modification will not seen by others. For the files we do not want them to see or modify, we just use bind mount to mount an empty folder on it or remount it as read only. To make this control solid, users inside a user container is forbidden to do mount and umount.

For devices, we also use bind mount to let users see the device which we want them to see. We also use bind mount to achieve read and write control. Mknod is forbidden inside user container because this is the job of the real root. Privileges about network are all dropped inside a user container.

Other capabilities related to system administration operations are all dropped. For example, CAP_SYS_ADMIN and CAP_SYS_MODULE may leave a door open to break into the kernel. CAP_SYS_BOOT may enable a user to do operations that affect others.

Last, CUE relies on Linux namespace to achieve isolation: pid namespace to isolate process pid; mount namespace to isolate changes to file system; utc namespace to isolate hostname, etc. There have been a lot of studies revealing vulnerabilities of Linux namespace. However, current namespace is quite mature and the attacking of certain namespace is beyond the scope of this paper [28-29].

To sum up, for security purpose, many of the capabilities are disabled for non-root users. After all, CUE is not meant to make non-root users become the true root. That is the work of traditional virtual machine. The spirit of CUE is to give users more freedom to make their own customization and improve privacy, to enhance the multi-user notion of current operating system.

IV. USE CASES AND COMPARISON

In this section we introduce some typical use case for CUE to highlight its contribution.

*A. Isolation and Customization*

The main purpose of CUE is to facilitate isolation and customization. For example, at first the pre-defined system environment has python of version 2.7. All users inside their user container see and use the python 2.7. Then a user (USERA) changes it to python 3.5. It is only the USERA sees and uses python 3.5 and others are still using 2.7. If the system administrator (root) updates the system python from 2.7 to 3.0, then all users except USERA see the update and uses python 3.0 from then. For USERA, he still sees and uses his customized version of 3.5. This design greatly gives users flexibility to make customization to fit their special needs while giving the right to the root user to maintain the base environment. For isolation, users are isolated mainly in terms of processes and files. Thus, they feel they are owning a computer exclusively except that they know there is a real root that is managing the computer for them underneath.

Moreover, CUE can help create different isolated environment for different projects that are hosted on one server. As we will show in experiment, the overhead is ignorable in terms of both time and space.

*B. Safe Update and Test of System Environment*

Another usage of CUE is for the real root to test the update the system environment safely. For example, without CUE, when the root wants to update several common libs, he just backups the old ones and replaces them with the new ones. After that the system may become unstable or may even crash. Much effort is engaged to recover the previous system state. With CUE, when the root wants to make update, he can just create a new user container. On the creation of the container, its environment is exactly the same with the host. Then the root can make update and test the environment well in the container. Once he has tested it well, he can merge the upper layer to the lower layer to make the update globally (see Figure 2). This process is much like making the update like a transaction, to be atomically.

*C. Integration into High Performance Clusters*

CUE is currently used in the prototype system (E prototype system) of our next generation of Tianhe super computer and we find this scenario is its best adoption.

Figure 3 shows the Common architecture of Tianhe supercomputer series. Users login into one of the login nodes (assigned automatically) to do program development and compiling. Once the program is ready to run, the user uses the cluster management system to allocate compute nodes and submit a job to run the program in the allocated compute nodes. To facilitate date exchange, a globally shared storage is mounted at every node.

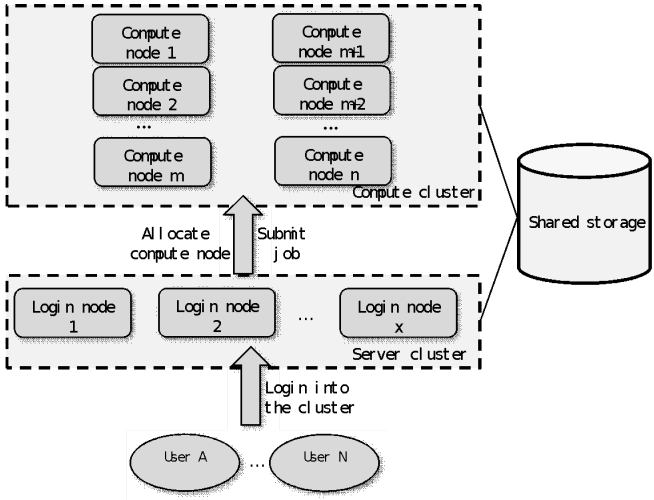

Fig. 3. Common architecture of Tianhe supercomputer series

CUE integrates into this scenario in the following two ways:

(1) Facilitate multi-user sharing a login node.

In high performance computer, often there are several users sharing a login node to do program development and compiling at the same time. Users are very sensitive of their privacy and they do not want others to know what they are running and to see their files. Thus, in this case CUE isolates users in terms of process and files.

Moreover, users may want to customize their environment. For example, they have a program that links to a special library. Without CUE, this is achieved by putting the special library in just somewhere the user can access, and the program must be recompiled to link to that place. With CUE, use can just put the library in the system default places (/lib or /usr/lib) and not affect others.

Finally, the root user can still manage the system environment for all user environments. CUE achieves a good balance in this scenario.

(2) Enable fast application deployment.

When a program needs to link to a special library, the original way is to first put the special library in the shared storage, and carefully set the link path in every allocated compute node so the compute node will find the library and run the program successfully. This deployment problem happens a lot. There have been a lot of studies trying to solve this problem for high performance computers. Shifter aims to introduce Docker's large ecosystem to high performance world. It transforms Docker image to make it suitable to run in high performance clusters. So users can build their Docker images at their own desktop and deploy it in easily in cluster. Singularity follows the same way and is more lightweight. It is just a packing program and isolates as less as possible. It is also compatible with Docker image and its goal is also to enable users to deploy the app they prepare ready in their own desktop.

However, high performance computers are highly customized, in terms of both hardware and software. The desktop on which users are building their Docker images may be quite different with the compute and login nodes in high performance clusters. For example, our Tianhe uses ARM ISA and Taihu-light uses Alpha ISA, while the majority of desktop and docker images are X86-based. Moreover, the software stack in high performance clusters are highly customized to fit its special hardware. Thus, the common scenario is that users need to develop and compile their program on the login node, not on their own desktop. In this scenario, the design of Shifter and Singularity is a bit redundant and could be optimized. For example, users use Shifter or Singularity command to pack their application well and send it to compute nodes to run. However, with CUE, we can achieve it easier.

With CUE on each login node, users actually login into their special user container, do customization there, and tested program well. Then when users submit the job to run a program, the job management system will just send the whole user container to the allocated compute nodes thus all customizations are synchronized. This synchronizing process is transparent to users and they do not bother to use Shifter or Singularity command to pack the application. Sending the whole user container to allocated compute node may sound heavyweight but the implementation is quite lightweight. We just put the upper layers of user containers in the shared storage. When a user logins into a login node, the upper layer is overlayed on the host "/" to form the user container. When a user submits a job, the corresponding upper layer is overlayed on the "/" of allocated compute nodes. Thus, the compute nodes sees all the customization the user has made and this enables a fast and transparent application deployment.

*D. Comparison*

To summary, we give a comparison of our CUE with related work.

TABLE I. COMPARESION WITH PREVIOUS WORK

| | OpenVZ VM | CUE | Docker | Shifter | Singularity |
|---|---|---|---|---|---|
| Main problem addressed | Virtualization of system environment or hardware. | Enhance multi-user isolation. User Environment isolation. | Application deployment. Application packaging. | | |
| Use mode | Transparent to users. Each VM is like black box to root. | Transparent to users. Root manages the base environment for all users. | Users use command to create and run containers. Uses themselves are not isolated nor privileged. | | |

| Isolation | High, almost everything isolated. | Moderate, process and file systems isolated. Devices and network shared. | High, by default isolate process, file systems, network, hostname, etc. | Low, for HPC use, does not isolate much. Rely on SLURM to do node-level isolation. | |
|---|---|---|---|---|---|
| GPU or Accelerator | Supported | Supported | Not support | Supported | On the way |
| Daemon | NA | No daemon | Daemon | No daemon | No daemon |
| SLURM-friendly | No | Yes | No | Yes | Yes |

## V. EXPERIMENT

The goal of CUE is to provide convenient features to multi-users while being lightweight and high performance. In this section we mainly test its overhead compared with bare metal (running directly without any container or virtualization).

The experimental platform is a login server of the Tianhe system. It is equipped with FT-1500A 16 cores CPU and 64GB of memory.

First, Figures 4-5 show the overhead of CUE compared with bare metal and docker on the benchmark NPB [30]. For all the benchmarks, we select relatively large scale and run every benchmark for ten times. The average result is presented.

We can see from the results that CUE introduces very little overhead (<1%). Running a process inside a user container is much like running it bare metal. The main overhead of containers is the overlay file system. However, as we use overlay file system in a limited way, we are able to limit the overhead. CUE just uses two layers of overlay. For most applications (especially high-performance computing application), they do not make modification to a large number of files in a short period, the overhead is very small. Docker introduces a little more overhead than CUE due to the overlay and namespace configuration: CUE uses less namespace than Docker.

As a same trend, Figures 6-7 show the overhead on the benchmark Sysbench [31] and the result shows that CUE introduces ignorable overhead.

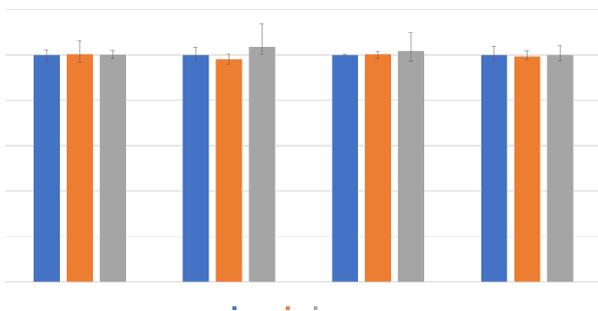

Fig. 4. Overhead of running NPB (MPI on 4 processes)

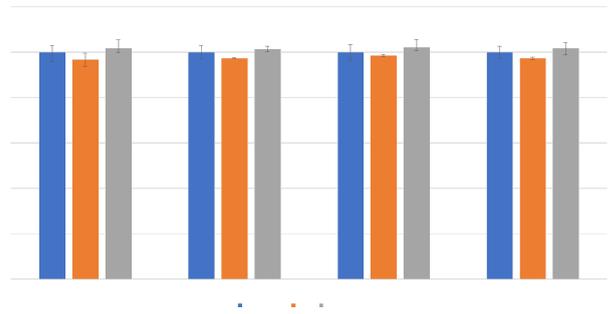

Fig. 5. Overhead of running NPB (serial)

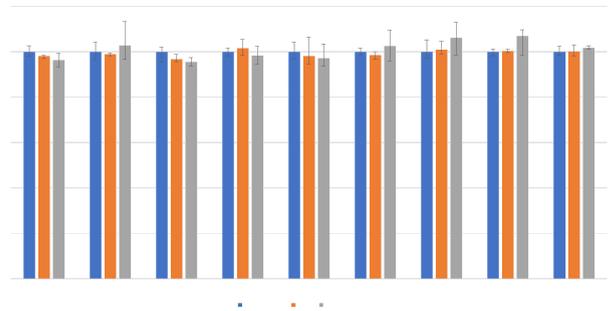

Fig. 6. Overhead of running CPU test of Sysbench

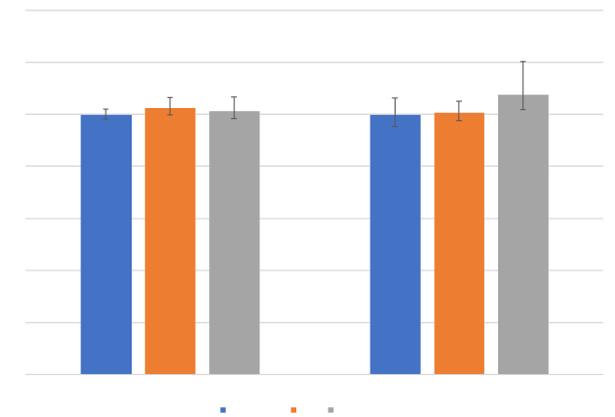

Fig. 7. Overhead of running MEM test of Sysbench

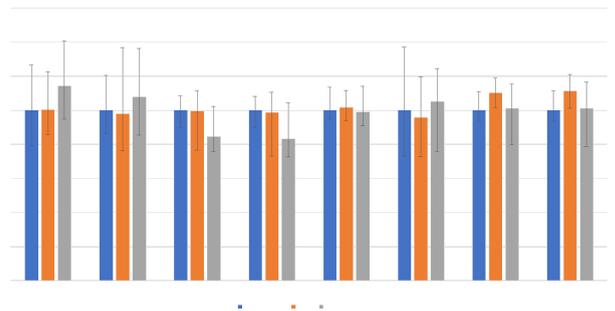

Fig. 8. Overhead of running Fio test

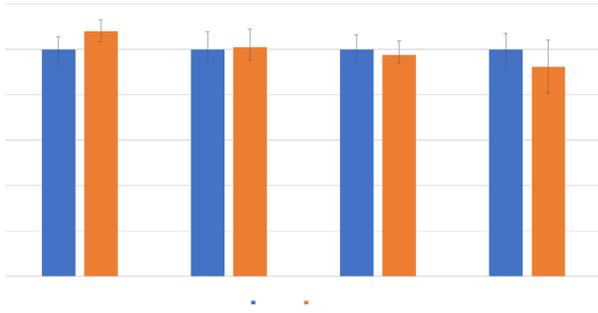

Fig. 9. Overhead of stress tests of file operations

Second, to show the overhead of file operations, we do Fio test and the result is shown in Figure 8. As we can see from the result, intense file operation is a main source of overhead for CUE, as well as Docker. Intense small files read or write would put great stress on overlay file system.

Last, we do stress test to show the overhead of file operation. Our stress tests are: (1) small file read and small file write: read and write a lot of small files (16B * 10000). (2) read and write a big file (1G). For file read, we read all the file content and redirect the content to /dev/null. For file write, we flip every bit (0 to 1 or 1 to 0) sequentially. In order to reveal the overhead of the layered file system on which CUE is based, before the test all files are generated on the host. That is to say, all files are in the lower layer at first. The result is shown in Figure 9.

We can see from Figure 9 that CUE introduces ignorable overhead for big file read and write. This is because the main overhead of the layered file system happens on managing file meta data. Thus, operating on small number of files would not hurt performance. This is also the typical scenario in high performance computing. Often HPC applications read or write to just a small number of big files. Second, reading a lot of small files also introduces little overhead, which just involves files searching in overlay file system. As CUE just uses 2 layers, this is not a performance problem. Last, the main overhead happens when writing to a lot of small files as shown in Figure 7. In this test CUE introduces about 4% overhead compared with bare-metal. This case is also the worst case for layered file system. We also argue that 4% overhead is acceptable and in most cases. In high performance computing, this seldom happens.

Last, we test the container startup time of CUE and compare it with Docker as the startup time is also an important figure of merit for containers [32-34]. In the experiment, we start 100 containers and record the time. The file systems are the same and we start bash inside the container. The result is shown in Table 1. The startup of CUE is much faster than Docker. This is mainly due to the lightweight implementation of CUE. CUE adopts a non-daemon model while Docker relies on a daemon to start the container. The non-daemon model makes it easy to integrate with the job management systems [35] in high performance clusters. Moreover, starting a new CUE causes negligible space overhead (less than 100kb) as we just overlap an empty folder on the "/" of the host. Thus, CUE provides a very lightweight container to use.

TABLE II. STARTUP TIME OF 100 CONTAINERS

|  | CUE | Docker |
|---|---|---|
| Startup time | 4.071s | 29.065s |

## VI. CONCLUSION

This paper proposes CUE, a Lightweight Container-based User Environment. CUE implements a new container notion and framework that stands in between application container and operating system container: user container. CUE gives users more flexibility to customize their environment, achieves privacy isolation, and makes system update easier and safer when multi-users are sharing a computer. Its goal is to optimize and enhance the multi-user notion of current operating system and being lightweight. It also facilitates application deployment in high performance clusters. It is currently deployed on NUDT's Tianhe E prototype supercomputer. Experiments show that CUE introduces negligible overhead. The startup time compared with Docker is more than one order of magnitude.

ACKNOWLEDGMENT

This work is supported by Tianhe Supercomputer Project 2018YFB0204301.